\documentclass[aps,pra,twocolumn,a4paper]{revtex4-1}
\usepackage{amsmath, amsthm, amssymb}
\usepackage{graphicx}
\usepackage{color}

\begin{document}

\title{Laser-beam scintillations for weak and moderate turbulence}

\author{R.~A.~Baskov}
\email{Email address: roman.baskov@iop.kiev.ua}
\affiliation{Institute of Physics of
the National Academy of Sciences of Ukraine,\\
 pr. Nauky 46, Kyiv-28, MSP 03028, Ukraine}

\author{O.~O.~Chumak}

\affiliation{Institute of Physics of
the National Academy of Sciences of Ukraine,\\
 pr. Nauky 46, Kyiv-28, MSP 03028, Ukraine}

\begin{abstract}
The scintillation index is obtained for the practically important range of weak and moderate atmospheric turbulence. To study this challenging range, the Boltzmann-Langevin kinetic equation, describing light propagation, is derived from first principles of quantum optics based on the technique of the photon distribution function (PDF) [G. P. Berman \textit{et al.}, Phys. Rev. A \textbf{74}, 013805 (2006)]. The paraxial approximation for laser beams reduces the collision integral for the PDF to a two-dimensional operator in the momentum space. Analytical solutions for the average value of PDF as well as for its fluctuating constituent are obtained using an iterative procedure. The calculated scintillation index is considerably greater than that obtained within the Rytov approximation even at moderate turbulence strength. The relevant explanation is proposed.

\end{abstract}

\maketitle

\section{Introduction}

Physics of light beam propagation in the Earth's atmosphere is of great interest for scientists and engineers, see \cite{Tatarskii1,Bar,Andrews,coro,fei}. This interest arises from applications in quantum and classical communications and remote sensing systems. The latest achievements in this field concern problems of quantum key distribution \cite{Capraro, Usenko}, propagation of entangled \cite{Ursin,Yin, Hosseinidehaj} and squeezed \cite{Peuntinger, Vasylyev} states, quantum nonlocality \cite{Semenov, Gumberidze}, quantum teleportation \cite{Ma, Ren}, tests of fundamental physical laws \cite{Rideout, Touboul}. In all these cases, random variations of the atmospheric refraction index distort the phase front of radiation causing intensity fluctuations  (scintillations), beam wandering and increasing beam spreading. Scintillations are the most severe problem which manifests itself in  a significant reduction of the signal-to-noise ratio (SNR) introducing  degradation of the performance of laser communication systems.

A laser beam in the Earth atmosphere is affected by turbulent eddies. Randomly distributed eddies stand for sources of local index-of-refraction fluctuations. There are numerous beam-eddies ``collisions" in the course of long-distance propagation. As a result, the radiation gradually acquires the Gaussian statistics. The scintillation index, $\sigma^2$, which is defined in classical optics as the inverse SNR, asymptotically approaches the level of $\sigma^2=1$. In this case, the intensity fluctuations are referred to as saturated \cite{Kravtsov}.

Scintillations are of importance for design of reliable classical and quantum optical communication systems \cite{Andrews2, Erven}, remote sensing systems \cite{Rino,Churnside}, and adaptive optics \cite{Ribak}. This field of research has also application in atmospheric physics, geophysics, ocean acoustics, planetary physics, and astronomy \cite{Tatarskii2}. The theoretical description of scintillation phenomena faces with the increasing computational complexity when one considers the parameter region of maximal optical beam intensity fluctuations. In order to overcome this problem several phenomenological and semi-phenomenological approaches were developed, which utilize the intensity distribution functions \cite{Jakeman}, phase screens \cite{Dashen-Wang}, turbulence spectrum approximation \cite{Marians}. The existing rigorous first-principles approaches, such as the method of smooth perturbations (Rytov approximation) \cite{Tatarskii1}, the Huygens-Kirchhoff method \cite{banakh79}, the path-integral method \cite{Das}, are applicable merely to the asymptotic regimes of weak and strong optical turbulences. At the same time, maximum scintillations lay in the region of moderate turbulence.

The range of moderate turbulence is the most challenging for rigorous theoretical study. First, the transition  from statistics of coherent laser beam to the Gaussian statistics lies just in this region.  Second, strong correlations of photon trajectories, which considerably enhance scintillations \cite{enha}, should also be taken into account here. A combined  effect of these important factors can lead to maximal scintillations. Such effect clearly manifests itself in various experiments where this maximum may considerably exceed the level of saturation \cite{Kravtsov,consortini,sedin}.

In the present paper we introduce for a first-principle approach for the description of weak and weak-to-moderate turbulence regimes, which remain the most challenging for the analysis. The method is based on the technique of the photon distribution function (PDF) \cite{Chu}, which is derived from the first principles of quantum optics.This method is applicable for an arbitrary quantum state of the light including coherent states, which describe laser-radiation fields.

The PDF is an operator-valued function, $\hat{f}(\mathbf{r},\mathbf{q})$, of the position $\mathbf{r}$ and the wave vector $\mathbf{q}$. It retains the concept of the Wigner function \cite{Wigner} such that the integration with respect to $\mathbf{q}$ or $\mathbf{r}$ results in the field intensity operator $\hat{I}(\mathbf{r})$ or the  photon-number operator $\hat{n}(\mathbf{q})$, respectively. The PDF can be found as a solution of the kinetic equation that accounts for random variations of the refractive index in the atmosphere. This approach has been originally introduced in the solid state physics (see, for example, Ref. \cite{chuZ}) and has also been successfully applied for a description of quantum radiation in waveguides \cite{sto,sto14}.

Application of the PDF method to the light propagation in the turbulent atmosphere has been considered in Refs. \cite{enha,Chu, Chumak wander}. It utilizes the approximation of the smoothly varying random force and is applicable only for restricted values of the turbulence parameters. In the present paper we derive a more general kinetic equation for the PDF introducing the collision integral and Langevin source of fluctuations. An approximate solution of this equation enables us to describe the beam characteristics beyond the Rytov approximation at the moderate range of turbulence, which was unreachable with the previous techniques.

To stress the significance of the present paper, it is worthwhile to recall the words of Dashen \cite{Das}. He considers ``the detailed behavior of the wave field at the boundaries between the unsaturated and saturated regimes" ``the remaining problem" in the physics of scintillation phenomena.  We hope that our paper as well as  the previous one \cite{enha} provide a deeper insight into physics and the theoretical description of this important region.

The rest of this paper is organized as follows. In Sec. \ref{sec:pdf} we give a brief review of the method of the PDF method. In Sec. \ref{sec:ble} and Appendix \ref{sec:appendix}, we explain the derivations of the collision integral and the corresponding Langevin source. In Sec. \ref{sec:scint_ind} and Appendix \ref{sec:appendix1}, we obtain an analytical formula for the scintillation index which is represented by a many-fold integral. In Sec. \ref{sec:discussion}, the results of numerical simulations are discussed. Concluding remarks are given in Sec. \ref{sec:conclusion}.

\section{Photon distribution function}
\label{sec:pdf}

The photon distribution function is defined in analogy to the widely used solid state physics distribution functions \cite{chuZ} (the distributions for electrons, phonons, etc). This function  is given by, see Ref. \cite{UJP},
\begin{equation}\label{1threee}
\hat{f}({\bf r},{\bf q},t)=\frac 1V\sum_{\bf k}e^{-i{\bf kr}}b^\dag_{{\bf
q}+ {\bf k}/2}b_{{\bf q}-{\bf k}/2},
\end{equation}
where $b^\dag_{\bf q}$ and $b_{\bf q}$ are bosonic creation and annihilation operators of photons  with the wave vector ${\bf q}$; $V\equiv L_xL_yL_z\equiv SL_z$ is the normalizing volume. We consider the laser beam propagating along the $z$ axis in the paraxial approximation. For this case the initial polarization of the beam  remains   almost undisturbed for a wide range of propagation distances, cf. Ref. \cite{stroh}.

The operator $\hat{f}({\bf r},{\bf q},t)$ describes the photon density in the phase  space (PDF in ${\bf r}{-}{\bf q}$ space). We consider the scenario with characteristic sizes of spatial inhomogeneities of the radiation field being much greater than the optical wavelength
$\lambda=(2\pi/q_0)$; here $q_0$ is the wave vector corresponding to the central frequency of the radiation, $\omega _0=cq_0$. In this case, it is reasonable to restrict the sum in Eq. (\ref{1threee}) by the range of small $k$, i.e. $k< k_0$ such that the inequality $k_0\ll q_0$ is satisfied.  At the same time the value of $k_0$ should be large enough to provide a desired accuracy for the description of the beam profile.

Evolution of the PDF $\hat{f}({\bf r},{\bf q},t)$ is governed by the Heisenberg equation
\begin{equation}\label{2five}
\partial_t \hat{f}({\bf r},{\bf q},t)=\frac 1{i\hbar }[\hat{f}({\bf r},{\bf
q},t),\hat{H}],
\end{equation}
where
\begin{equation}\label{3six}
\hat{H}=\sum_{\bf q}\hbar\omega_{\bf q}b^\dag_{\bf q}b_{\bf q}-\sum_{\bf
q,k}\hbar\omega_{\bf q}n_{\bf k}b^\dag_{\bf q}b_{\bf q+k}
\end{equation}
is the Hamiltonian of photons in a medium with a fluctuating refraction index $n({\bf r})=1+\delta n ({\bf r})$, where $\delta n ({\bf r})$ stands for fluctuating part representing atmospheric inhomogeneity. The quantities
$\hbar\omega_{\bf q}=\hbar cq$ and ${\bf c_q}=\frac{\partial
\omega_{\bf q}}{\partial{\bf q}}$ are the photon energy
and photon velocity in vacuum, $n_{\bf k}$ is the Fourier transform of the fluctuating refraction index
 $\delta n({\bf r})$ is defined by
\begin{equation}\label{two}
n_{\bf k}=\frac 1V\int dVe^{i{\bf kr}}\delta n({\bf r}).
\end{equation}

By substituting the Hamiltonian \eqref{3six} into Eq.  \eqref{2five}, the latter is rewritten as
\begin{eqnarray}\label{5seven}
\partial_t \hat{f}({\bf r},{\bf q},t)+{\bf c_q}\cdot\partial_{\bf r}\hat{f}({\bf r},{\bf q},t)-i\frac{\omega _0}{V}\sum_{{\bf k},{\bf k'}}e^{-i{\bf k\cdot r}}n_{{\bf k}^\prime}\nonumber\\
\times\big[b^\dag _{\bf q+ \frac {k}2}b_{\bf q-\frac{k}{2}+k^\prime}-b^\dag _{\bf q+ \frac {k}2-k^\prime}b_{\bf q-\frac{k}{2}} \big]=0.
\end{eqnarray}
The first two terms in the left-hand side describe free-space propagation of a laser beam and the last term arises from atmospheric inhomogeneity. The latter can be replaced by
${\bf F}({\bf r})\cdot\partial_{\bf q}\hat{f}({\bf r},{\bf q},t)$ if three components of the turbulence wave vectors ${\bf k}^\prime$ are much smaller than the corresponding characteristic values of ${\bf q}$, i.e. we can express the difference of functions in square brackets in Eq. (\ref{5seven}) by the corresponding derivative. The quantity ${\bf F}({\bf r})=\omega _0\partial_{\bf r}n({\bf r})$  is interpreted as a random force produced by atmospheric vortices \cite{Chu}. With this force  Eq. (\ref{5seven}) takes the form of the kinetic equation
\begin{equation}\label{6seven1}
\partial_t \hat{f}({\bf r},{\bf q},t)+{\bf c_q}\cdot\partial_{\bf r}\hat{f}({\bf r},{\bf q},t)+
{\bf F}({\bf r})\cdot\partial_{\bf q}\hat{f}({\bf r},{\bf q},t)=0.
\end{equation}
This equation resembles the collisionless Boltzmann equation with a smoothly varying momentum-independent force {\bf F}({\bf r}) acting on point-like particles.

The technique of the PDF (see Refs. \cite{Chu, enha, ChuSingle, Chumak wander, ChuPhase, sto, sto14}) is convenient for obtaining average parameters of the beam as well as for the description of wave-field  fluctuations. The distribution function describes the photon density in the configuration-momentum phase space. A solution of the kinetic equation (\ref{6seven1}) with a smoothly varying fluctuation force has been obtained in Refs. \cite{Chu, enha, ChuSingle, Chumak wander, ChuPhase}. This simplified physical picture is justified only if the photon momentum \cite{name}, $\bf{q}$, is much greater than the inverse size of eddies. All components of ${\bf q}$ should obey this requirement. In the paraxial approximation, the perpendicular components of photon wave vector, $\bf q_\bot$, increase with the propagation time $t$ as $t^{1/2}$  \cite{Chu} and the beam inevitably reaches  the region of saturated scintillations if $t{\rightarrow}\infty$. This indicates that Refs. \cite{Chu, chuZ, Chumak wander, ChuSingle, ChuPhase} consider the strong turbulence regime, including the limiting case of saturation, rather than the regime of a weak turbulence. The range, where the random force can be considered as smoothly varying function, extends towards smaller distances if the phase diffuser is used. The reason for this is that the phase diffuser increases  the characteristic values of ${{\bf q}_\bot}$; see Refs. \cite{Ban55,Ban54,Chu} for more details.

\section{Boltzmann-Langevin equation}
\label{sec:ble}

The scheme for derivation of the kinetic equation (\ref{6seven1}), outlined in previous section, can be justified when all components of photon wave vector ${\bf q}$ are sufficiently large. The corresponding situation occurs for long-distance propagation or strong turbulence (see, for example, Sec. VI in \cite{Chu}). It should be emphasized that it is just the case when the direct computer simulation of beam propagation becomes problematic \cite{Gorshkov}.  In what follows, we describe more general approach which is free from this undesirable restriction.

In the kinetic equation (\ref{5seven}) the last left-hand-side term describes process of photon ``collisions " with atmospheric inhomogeneities. The amplitude of this process is determined by $n_{\bf k^\prime}$ which is a random quantity with $\langle n_{\bf k^\prime}\rangle =0$. Two operators in square brackets of Eq. (\ref{5seven}) also depend on $ {\bf k}^\prime$. Their explicit dependence on the random refraction index can be obtained from the Heisenberg equations. One of them is given by
\begin{eqnarray}\label{7nine}
\!\left\{\partial_t - i\left(\omega_{\bf q+\frac{k}{2}}-\omega_{\bf q-\frac{k}{2}+{k^\prime}}\right) \right\}b^\dag_{\bf q+\frac{k}{2}}b_{\bf q-\frac{k}{2}+k^\prime}=\nonumber\\
i\omega_0\sum_{{\bf {k}^{\prime\prime}}}n_{\bf k^{\prime\prime}}\bigg[b^\dag_{\bf q+\frac{k}{2}} b_{\bf q-\frac{k}{2}+k^{\prime}+{k}^{\prime\prime}}-b^\dag_{\bf q+\frac{k}{2}-{k}^{\prime\prime}}b_{\bf q-\frac{k}{2}+k^\prime}\bigg].
\end{eqnarray}
Its solution can be written as
\begin{eqnarray}\label{8ten}
b^\dag_{\bf q+\frac{k}{2}}&&b_{\bf q-\frac{k}{2}+k^\prime}\bigg|_t=
e^{i(\omega_{\bf q+\frac{k}{2}}-\omega_{\bf q-\frac{k}{2}+{k^\prime}})(t-t_0)}\left(b^\dag_{\bf q+\frac{k}{2}}b_{\bf q-\frac{k}{2}+{k^\prime}}\right)\bigg|_{t_0}\nonumber\\
&&+i\omega_0\sum_{\bf {k}^{\prime\prime}}\int\limits_{t_0}^{t}dt^\prime e^{i(\omega_{\bf q+\frac{k}{2}}-\omega_{\bf q-\frac{k}{2}+{k'}})\left(t-t'\right)}n_{\bf {k}^{\prime\prime}}\nonumber\\
&&\times\bigg(b^\dag_{\bf q+\frac{k}{2}}b_{\bf q-\frac{k}{2}+{k'}+{k}^{\prime\prime}}-b^\dag_{\bf q+\frac{k}{2}-{k}^{\prime\prime}}b_{\bf q-\frac{k}{2}+{k'}}\bigg)\bigg|_{t^\prime},
\end{eqnarray}
where the subscripts $t_0$ and $t^\prime$ indicate the dependence of the corresponding operators on time.

In Eq. (\ref{8ten}) the interval $t-t_0$ is chosen to be large compared with the photon-eddy interaction time $\pi/ck^\prime$ and sufficiently short  compared with the relaxation time $1/\nu$ caused by these interactions:
\begin{equation}\label{9tenprime}
\pi/ck^\prime \ll t- t_0\ll1/\nu .
\end{equation}
 Here $\nu$ is the collision frequency and the quantity $1/k^\prime$ describes the characteristic length of atmospheric inhomogeneities. In other words, the time hierarchy  (\ref{9tenprime})  means that the duration of photon interaction with  scatterers is much shorter than the time of free flight. This is a typical criterion ensuring applicability of the Boltzmann  equation for the description of many-particle systems (see, for example, Ref. \cite{chuZ}) .

Substituting Eq. (\ref{8ten}) and a similar solution for the operator $b^\dag _{\bf q+ \frac {k}2-k^\prime}b_{\bf q-\frac{k}{2}}\big|_t$ into Eq. (\ref{7nine}) we obtain the kinetic equation for  $\hat{f}({\bf r},{\bf q},t)$

\begin{equation}\label{10eleven}
\partial_t \hat{f}({\bf r},{\bf {q}},t)+{\bf c_q}\cdot\partial_{\bf r}\hat{f}({\bf r},{\bf q},t)=
\hat{K}({\bf r},{\bf q},t)-
\hat{\nu}_{\bf q}\big \{ \hat{f}({\bf r},{\bf q},t)\}  ,
\end{equation}
where
\begin{widetext}
\begin{equation}\label{11elevven}
\hat{K}({\bf r},{\bf q},t){=}\frac{i\omega_0}{V}\sum_{{\bf k,k^\prime}}e^{-i{\bf k}\cdot{\bf r}}n_{\bf k^\prime}\big[e^{i(\omega_{\bf q+\frac{k}{2}}-\omega_{\bf q-\frac{k}{2}+k^\prime})(t-t_0)}\big(b^\dag_{\bf q{+}\frac{k}{2}}b_{\bf q{-}\frac{k}{2}{+}k'}\big){}\big|_{t=t_0}{-}
 e^{i(\omega_{\bf q+\frac{k}{2}-k^\prime}-\omega_{\bf q-\frac{k}{2}})(t-t_0)}\big(b^\dag_{\bf q{+}\frac{k}{2}{-}k^\prime}b_{\bf q{-}\frac{k}{2}}\big)\big|_{t=t_0}\big],
\end{equation}
\begin{equation}\label{12twelwe}
\hat{\nu}_{\bf q}\big \{ \hat{f}({\bf r},{\bf q},t)\}=\frac{2\pi\omega_{0}^{2}}{c}\int d{\bf k'_{\bot}}\psi({\bf k'_{\bot}})\big(\hat{f}({\bf r},{\bf q},t)-\hat{f}({\bf r},{\bf q+k'_{\bot}},t)\big).
\end{equation}
\end{widetext}
The notation $(_\bot)$ indicates components of the corresponding vector perpendicular to the $z$-axis, and  $\psi ({\bf k'_{\bot}})=\frac V{(2\pi)^3}\langle|n_{\bf k'_{\bot}}|^2\rangle$. The value of  $\psi ({\bf k})$ is given by the von Karman formula
\begin{equation}\label{13twelwwe}
\psi ({\bf k})=0.033C_n^2\frac {\exp(-(kl_0/2\pi
	)^2)}{(k^2+L_0^{-2})^{11/6}},
\end{equation}
where the structure constant $C_n^2$ describes the strength of the index-of-refraction fluctuations, whereas  $L_0$ and $l_0$ are usually referred to as the outer and inner radii of the turbulent eddies, respectively. These radii restrict a range of characteristic values of $\bf k'_{\bot}$. In atmospheric turbulence, $L_0$ may range from 1 to 100 m, and $l_0$ is on the order of few millimeters.
It is seen from Eqs. (\ref{10eleven})-(\ref{13twelwwe}) that the random quantity $\hat{K}({\bf r},{\bf q},t)$ linearly depends on $n_{\bf k^\prime}$, while $\hat{\nu}_{\bf q}$ depends only on a regular variable $\langle|n_{\bf k'_{\bot}}|^2\rangle$. The contribution of fluctuating part of $n_{\bf{k}'} n_{\bf{k}''}$ can be neglected (for more details see Appendix \ref{sec:appendix}).

The linear inhomogeneous equation (\ref{10eleven}) governs the evolution of photon distribution in the phase space. The term $\hat{\nu}_{\bf q}\big \{ \hat{f}({\bf r},{\bf q},t)\}$ describes dissipation of the distribution function caused by randomization of the photon wave vector ${\bf q_\bot}$. The term ``dissipation" does not mean here that the total number of photons decreases. Actually, after summing up the collision term (\ref{12twelwe}) over $\bf q$ we get zero, which indicates that  the photon number is conserved.
The collision frequency $\nu$ can be estimated by $\frac{2\pi\omega_0^2}c\psi (k'_{\bot})k'^2_\bot$, where $ k'_\bot$ is the characteristic value of the momentum transfer.

The  Langevin source of fluctuations in Eq. (\ref{10eleven}) is represented  by $\hat{K}({\bf r},{\bf q},t)$. Random photon-eddy ``collisions" (see Refs. \cite{chuZ} and \cite{Kogan}) generate the Langevin source. Within the time interval, restricted by the inequality (\ref{9tenprime}), the constituents in the right-hand side of Eq. (\ref{11elevven}) have a simple oscillating dependence on time.  Due to this favorable circumstance, the calculation of two-time correlation function $\langle \hat{K}({\bf r},{\bf q},t)\hat{K}({\bf r}^\prime,{\bf q}^\prime,t^\prime)\rangle$ reduces to obtaining the average value of the operator products defined at the same time, $t_0$.   The source vanishes after averaging of Eq. (\ref{10eleven}). Then the remaining homogeneous equation for $\langle \hat{f}({\bf r},{\bf q},t)\rangle$ can be used for obtaining parameters of the beam at any distances.[In what follows, we use $f({\bf r},{\bf q},t)$ notation for $\langle \hat{f}({\bf r},{\bf q},t)\rangle$]. For long-distance propagation, where
\begin{equation}\label{14twe}
 q_{\bot}\gg k'_{\bot},
\end{equation}
the collision integral reduces to the differential form
\begin{equation}\label{15twell}
\hat{\nu}_{\bf q}\big \{ \hat{f}({\bf r},{\bf q},t)\}=-\frac{\pi\omega_{0}^{2}}{c}\int d{\bf k'_{\bot}}\psi({\bf k'_{\bot}})\bigg(\frac \partial{\partial{\bf q}} {\bf k^\prime}_\bot \bigg)^2\hat{f}({\bf r},{\bf q},t),
\end{equation}
which describes a diffusion-like motion in the wave vector space.

The kinetic equation with $\hat{K}({\bf r},{\bf q},t)=0$ and the collision term, which is similar to (\ref{15twell}), was used in Refs. \cite{Yang51} and \cite{berman95} to investigate the propagation of relativistic charged particles through an inhomogeneous medium (for example, through a foil). The similarity arises from equivalence of the small-scattering-angle approximation, used in Refs. \cite{Yang51}, \cite{berman95}, and  the paraxial approximation, used in this paper. Although the linear energy-momentum relationship holds for both the photons and ultrarelativistic particles, the microscopic scattering mechanisms are different for those cases.

\section{Scintillation index}
\label{sec:scint_ind}
Equation (\ref{10eleven}) can be used to study the effect of photon multiple scattering on their distribution in the phase space. Summation of $\hat{f}({\bf r},{\bf q},t)$ over ${\bf q}$ results in a spatio-temporal photon distribution
\begin{equation} \label{16twel}
\hat{I}({\bf r},t)=\sum_{\bf q}\hat{f}({\bf r},{\bf q},t),
 \end{equation}
which includes an average value, $\langle \hat{I}({\bf r},t)\rangle\equiv I({\bf r},t) $, and fluctuations, $\delta \hat{I}({\bf r},t)$,
\begin{eqnarray} \label{17svnt}
 \hat{I}({\bf r},t)&{=}& I({\bf r},t)+\delta \hat{I}({\bf r},t)\nonumber\\
 &{=}&\sum_{\bf q} f({\bf r},{\bf q},t)+\sum_{\bf q}\delta \hat{f}({\bf r},{\bf q},t),
 \end{eqnarray}
where $\delta \hat{f}({\bf r},{\bf q},t)=\hat{f}({\bf r},{\bf q},t)-f({\bf r},{\bf q},t)$.

To obtain $I({\bf r},t)$, one needs to solve averaged Eq. (\ref{10eleven}), accounting for the boundary conditions at the aperture plain and using  $\langle \hat{K}({\bf r},{\bf q},t)\rangle=0$.

The scintillation index is defined by
\begin{equation}\label{18fifte}
\sigma^2=\frac {\langle : \delta \hat{I}^2({\bf r}):\rangle}{ I({\bf r})^2}=\frac {\langle : \hat{I}^2({\bf r}):\rangle-I({\bf
r})^2}{ I({\bf r})^2},
   \end{equation}
where the symbol $\{:..:\}$ means the normal ordering of the creation and annihilation operators. The definition (\ref{18fifte}) does not include contribution of shot noise. This noise enters the fluctuations of the detector counts and tends to be important in problems of quantum optics. The shot-noise term is linear in the photon density. It can be easily excluded from experimental data to facilitate the comparison with the theoretical calculation.

Calculation of Eq. (\ref{18fifte}) is more intricate. It follows from Eqs. (\ref{17svnt}) and (\ref{18fifte}) that $\sigma^2$ is a quadratic form of PDF fluctuations,  $\langle\delta \hat{f}({\bf r},{\bf q},t)\delta \hat{f}({\bf r}^\prime,{\bf q}^\prime,t^\prime)\rangle$. Hence, the calculation of $\sigma^2$ is possible if the correlation function  of photon distributions is known. To simplify the problem, we use an approximate iterative scheme.

\subsection{First order approximation}
The approximation is based on the assumption that close to the transmitter aperture the collision term does not perturb significantly PDF and can be omitted. In this case, the average value of PDF satisfies the equation
\begin{equation}\label{19ninn}
(\partial_t+{\bf c_q}\cdot\partial_{\bf r})f_0({\bf r},{\bf q},t)=0.
\end{equation}
 The fluctuating part of  $\delta \hat{f}({\bf r},{\bf q},t)$  is governed by the similar equation supplemented with the Langevin source $\hat{K}$
 \begin{equation}\label{20nint}
 (\partial_t+{\bf c_q}\cdot\partial_{\bf r})\delta \hat{f}({\bf r},{\bf q},t)=\hat{K}({\bf r},{\bf q},t).
 \end{equation}

Equations (\ref{19ninn}) and (\ref{20nint}) follow from Eq. (\ref{10eleven}) after replacing $\hat{f}$ by $f_0+\delta \hat{f}$. The Langevin source linearly depends on $n_{{\bf k}_\bot}$ while the neglected collision integral is quadratic in $n_{{\bf k}_\bot}$. Therefore, Eqs. (\ref{19ninn}) and (\ref{20nint}) can be interpreted as the lowest-order expansions of Eq. (\ref{10eleven})  in powers of $n_{{\bf k}_\bot} $.
 The general solution of Eq. (\ref{20nint}) is represented by two terms
  \[\delta \hat{f}({\bf r},{\bf q},t)=\delta \hat{f}_0({\bf r_q}(t'),{\bf q},t')|_{t'=0}+\delta \hat{f}_1({\bf r,q},t),\]
 where  ${\bf r_q}(t^\prime)={\bf r}-{\bf c_q}(t-t')$ and
\begin{equation}\label{21twnt}
  \delta  \hat{f}_1({\bf r,q},t)=\int\limits_{0}^{t}dt'\hat{K}({\bf r_q}(t'),{\bf q},t').
\end{equation}
We consider the aperture plane as a starting points of photon trajectories (at $t'=0$). The  paraxial approximation imposes a set of restrictions on the wave-vectors: $q_z{\sim}q_0{\gg}q_\bot,k_\bot,k_\bot^\prime$. Then $z_{\bf q}(t^\prime =0)=z-ct=0$.

 The term, $\delta \hat{f}_0({\bf r_q}(t'),{\bf q},t')|_{t'=0}$,  describes the evolution of PDF fluctuations in vacuum. In what follows, we  neglect fluctuations of the incident light. In this case
 $\delta \hat{f}_0({\bf r_q}(t'),{\bf q},t')|_{t'=0}=0$ and only the  term, $\delta \hat{f}_1({\bf r,q},t)$, is responsible for the non-zero amount of the scintillation index, $\sigma^2$, at small propagation time $t$. It is given by
\begin{widetext}
  \begin{equation}\label{22twnt1}
\sigma^2=\frac {\sum_{\bf q, q'}\langle:\delta \hat{f}({\bf r},{\bf q},t)\delta \hat{f}({\bf r},{\bf q}^\prime,t):\rangle }{ (\sum_{\bf q}f_0({\bf r},{\bf q},t))^2}=\frac {\sum_{\bf q, q'}\int\limits_{0}^{t}\int\limits_{0}^{t}dt'dt''\langle :\hat{K}({\bf r_q}(t'),{\bf q},t')\hat{K}({\bf r_{q^\prime}}(t''),{\bf q'},t''):\rangle }{ (\sum_{\bf q}f_0({\bf r},{\bf q},t))^2},
\end{equation}
where
\begin{equation}\label{23twntx}
 f_0({\bf r},{\bf q},t)=  f_0({\bf r_q}(0),{\bf q},0),\quad
\sum_{\bf q} f_0({\bf r},{\bf q},t) \equiv I_0({\bf r},t)={1\over V}  \sum_{\bf q,k}e^{-i{\bf k}({\bf r-c_q}t)}\langle b^\dag_{{\bf q+\frac k2}} b_{{\bf q-\frac k2}}\rangle |_{t=0}.
\end{equation}
\end{widetext}
The first equation in (\ref{23twntx}) means that the  left-hand-side term satisfies both the collisionless kinetic equation (\ref{19ninn}) and the boundary conditions at the aperture.
The value of $ I_0({\bf r},t)$ is equal to photon density in the absence of turbulence.

The numerator in the right-hand side of Eq. (\ref{22twnt1}) can be calculated using the explicit term (\ref{11elevven}) for $\hat{K}({\bf r},{\bf q},t)$ and meeting boundary conditions (see App. \ref{sec:appendix1}).
Then the scintillation index linearly depends on $\langle|n_{\bf k_\bot}|^{2}\rangle$ and reduces to
\begin{equation}\label{35twnt66}
\sigma ^2=\sigma _1^2L(z,\rho_0,\rho_1),
\end{equation}
where $\sigma _1^2=1.23C_n^2q_0^{7/6}z^{11/6}$ is the Rytov variance,
 $\rho _{0,1}^2={r_{0,1}^2q_0}/z $, $r_0$ is initial radius of the beam, $r^2_1=r_0^2/(1+2r_0 ^2\lambda _c^{-2})$, the quantity $\lambda _c$ describes the effect of the phase diffuser, and  $L(z,\rho_0,\rho_1)$ is the double integral
\begin{equation}\label{36twnt7}
L(z,\rho_0,\rho_1)=4.24\int\limits _0^1d\tau \int\limits _0^\infty d\chi \chi^{-8/3}\exp\Bigg\{-\chi^2\Bigg[\frac
{q_0l_0^2}{4\pi ^2z}+
\end{equation}
\[
\tau ^2\frac {\rho_0^2+\rho_1^2}{4+\rho_0^2\rho_1^2}\Bigg]\Bigg\} \sin^2\Bigg(\frac {\tau \chi^2}2-\frac {2\tau ^2\chi^2}{4+\rho _0^2\rho _1^2}\Bigg).\]
 Equations (\ref{35twnt66}) and (\ref{36twnt7}) were  derived   in \cite{Chu} using a  different approach. It follows from these equations that in the limit of large initial radius of beam aperture ($\rho_0,\rho_1{\rightarrow}\,\infty$)  and infinitely small inner scale of turbulence ($l_0{\rightarrow}\,0$), we have the result of Rytov theory ($\sigma ^2=\sigma _1^2$) because $L{\rightarrow}\,1$.

\subsection{Collision term in average intensity}

The numerator as well as the denominator in Eq. ($\ref{22twnt1}$) are derived using only first non-vanishing iterative terms. Extension of the theory towards a moderate turbulence requires accounting for the collision term $-\hat{\nu}\big \{ \hat{f}({\bf r},{\bf q},t)\}$. Following the iterative procedure, we  substitute the approximate value of PDF, given by Eq. (\ref{23twntx}), into the collision term of Eq. (\ref{10eleven}). Then the right-hand side of Eq. (\ref{10eleven}) is considered as a known function. After averaging the modified equation, we obtain
\begin{equation}\label{24elevenx}
(\partial_t +{\bf c_q}\cdot\partial_{\bf r}) f_1({\bf r},{\bf q},t)=-\hat{\nu}_{\bf q}\big \{  f_0({\bf r},{\bf q},t)\},
\end{equation}
where $ f_1$ is the first non-vanishing term generated by the collision integral. Solution of Eq. (\ref{24elevenx}), obeying zero-value boundary conditions, is given by
\begin{equation}\label{25therty}
 f_1({\bf r},{\bf q},t)=-\int\limits _0^tdt^\prime\hat{\nu}_{\bf q}\{ f_0({\bf r_q}(t^\prime),{\bf q},t    ^\prime)\}.
\end{equation}
The  contribution of $ f_1({\bf r},{\bf q},t)$ into the total photon density is given by
\begin{eqnarray}\label{26twnt}
 I_1({\bf r},t) \equiv \sum_{\bf q} f_1&&({\bf r},{\bf q},t)=-\frac{\omega_0^2t}{cS}\sum_{\bf q,k,k_\bot^\prime}\langle|n_{\bf k_\bot^\prime}|^2\rangle e^{-i{\bf k}({\bf r-c_q}t)}\nonumber\\
&&\times\bigg[1-\frac {\sin({\bf kc_{k_\bot ^\prime} }t)}{{\bf kc_{k_\bot ^\prime} }t}\bigg]\langle b^\dag_{{\bf q+\frac k2}} b_{{\bf q-\frac k2}}\rangle|_{t=0}.
\end{eqnarray}
Equation (\ref{26twnt}) accounts for the beam broadening caused by atmospheric eddies. Averaging of each factor in the sum can be performed independently because of the absence of correlations between the source fluctuations and the refractive index fluctuations.

Two quantities, $ I_0({\bf r},t)$ and $ I_1({\bf r},t)$, are zeroth- and first-order terms of the development of average photon density in powers of $\langle|n_{\bf k_\bot}|^2\rangle$, respectively.  

\subsection{Second order $\delta \hat{f}_{2}$ and combined effect of fluctuations  $\delta \hat{f}_{1}{\cdot}\delta \hat{f}_{2}$}

The second iterative term for fluctuations of PDF, $\delta \hat{f}_2$, obeys the equation
\begin{equation}\label{27ninty}
\partial_t \delta \hat{f}_2({\bf r},{\bf q},t)+{\bf c_q}\cdot\partial_{\bf r}\delta \hat{f}_2({\bf r},{\bf q},t)=-{\hat \nu}_{\bf q}\{\delta \hat{f}_1({\bf r_q},{\bf q},t)\},
\end{equation}
where the function $\delta \hat{f}_1$, given by Eq. (\ref{21twnt}),  enters the collision term. Solution of Eq. (\ref{27ninty}) is
\begin{equation}\label{28twnt9y}
\delta \hat{f}_2({\bf r},{\bf q},t)=-\int\limits_{0}^{t}dt'\hat{\nu}_{\bf q} \{\delta \hat{f}_1({\bf r_q}(t^\prime),{\bf q},t^\prime)\},
\end{equation}
were the explicit form of the collision integral is given by
\begin{eqnarray}\label{29twnt9yy}
&{{\hat \nu}_{\bf q}\{\delta \hat{f}_1({\bf r_q}(t^\prime),{\bf q},t')\} =\frac{L_z\omega_0^2}{c} \sum\limits_{{\bf k'_\bot}} \langle |n_{\bf k'_\bot}|^2\rangle}\nonumber\\
&\times\big[\delta \hat{f}_1({\bf r_q(t^\prime)},{\bf q},t')-\delta \hat{f}_1({\bf r_q(t^\prime)},{\bf q+k'_\bot},t')\big ].
\end{eqnarray}
 To proceed, let us consider a combined effect of fluctuations  $\delta \hat{f}_{1,2}({\bf r},{\bf q},t)$ on $\sigma^2$.  Contributions of $\delta \hat{f}_{1,2}$ into the photon density are given by $\sum\limits_{\bf q}(\delta \hat{f}_1({\bf r},{\bf q},t)+\delta \hat{f}_2({\bf r},{\bf q},t))$. This sum includes linear and cubic in $n_{\bf{k_\bot}}$ terms.  The average square of this sum includes the  term

 \begin{eqnarray}\label{37p}
 	&\sum_{{\bf q,q}_1}&\langle\delta\hat{f}_1({\bf r},{\bf q},t){\cdot}{\delta} \hat{f}_2({\bf r},{{\bf q}_1},t)+\delta \hat{f}_2({\bf r},{\bf q},t)\cdot\delta \hat{f}_1({\bf r},{{\bf q}_1},t)\rangle\nonumber\\
 &=&{2\sum_{{\bf q,q}_1}\langle\delta \hat{f}_1({\bf r},{\bf q},t)\cdot\delta \hat{f}_2({\bf r},{{\bf q}_1},t)\rangle}
 \end{eqnarray}
which is quadratic in $\langle|n_{{\bf k}_\bot}|^2 \rangle$. For obtaining  $\sigma^2$, we use this term and neglect terms of order $O(\langle|n_{\bf k_\perp}|^2\rangle^3)$. Then using Eqs. (\ref{21twnt}) and (\ref{28twnt9y}) we obtain the explicit expression for Eq. (\ref{37p}). It is given by

 \begin{eqnarray}\label{38p}
 2&\sum\limits_{{\bf q,q}_1}&\langle\delta \hat{f}_1({\bf r},{\bf q},t)\cdot\delta \hat{f}_2({\bf r},{{\bf q}_1},t)\rangle\nonumber\\
 &=&\frac{2\omega_0^4}{c^2S^2}\sum_{\substack{{\bf q,k,k^\prime} \\ {\bf q_1,k_1,k^{\prime\prime}}}}\langle|n_{{\bf k}^\prime}|^2\rangle\langle| n_{{\bf k}^{\prime\prime}}|^2\rangle\int\limits _0^td\tau\nonumber\\
 &\times&\int\limits _\tau^td\tau_1 e^{-i{\bf k\cdot(r-c_q\tau)}-i{\bf k}_1\cdot({\bf r-c_q}_1 \tau_1)}\nonumber\\
 &\times&\big[1-e^{-i{{\bf k\cdot c_{k^{\prime}}}\tau)}}\big]\big[1-e^{i{{\bf k\cdot c_{k^{\prime\prime}}}\tau_1)}}\big]\big[1-e^{-i{{\bf k_1\cdot c_{k^{\prime\prime}}}\tau_1)}}\big]\nonumber\\
 &\times&\big\langle b^\dag_{{\bf q+k}/2} b^\dag_{{\bf q_1+k_1}/2} b_{{\bf q-k}/2+{\bf k^{\prime\prime}}} b_{{\bf q_1-k_1}/2-{\bf k^{\prime\prime}}}\big\rangle\big|_{t-\tau_1},
 \end{eqnarray}

where the operators in the angle brackets depend on time as in the absence of turbulence.

 The summation in Eq. (\ref{38p}) runs over components of vectors ${\bf q},{\bf q}_1,{\bf k},{\bf k}_1,{\bf k}^\prime,{\bf k}^{\prime\prime}$ which are perpendicular to the $z$-axis  (the labels ($_\bot$) are omitted for brevity). Parallel to the $z$-axis components are given by
 \begin{equation}\label{39p}
 q_z=q_{1z}=q_0,\quad k_z=k_{1z}=k^\prime_z=k^{\prime\prime}_z=0.
 \end{equation}
  The relations (\ref{39p}) can be derived  from Eq. (\ref{32twnt5xx}).

 The conditions  $k_z^\prime =k_{z}^{\prime\prime}=0$ are consistent with the Markov approximation \cite{Tatarskii1}, \cite{Fante1} (not used here!) in which the index-of-refraction fluctuations, $\delta n({\bf r})$, are assumed to be delta-function correlated in the direction of propagation:
  \[\langle\delta n({\bf r}_\bot,z)\delta n({\bf r}^\prime_\bot,z^\prime)\rangle\sim  \delta (z-z^\prime).\]
In this case, the turbulent eddies look like flat disks oriented normally to the propagation path. At first sight, this representation of the correlation function seems unrealistic because the atmosphere is assumed to be statistically homogeneous and isotropic. The paradox is explained by the effect of relativistic length contraction (Lorentz contraction) of moving objects. The relative motion of the atmosphere towards photons results in a zero value of correlation length in the direction of motion.

The effect of turbulence comes only from ``diagonal" components  $\langle|n_{{\bf k}^\prime_\bot}|^2\rangle$ and $\langle|n_{{\bf k}^{\prime\prime}_\bot}|^2\rangle$ of the correlation function. As before, this is the result of statistical homogeneity of the turbulent atmosphere.

The final result of this Section is represented by
\begin{equation}\label{40p}
 \sigma^2=\frac{\sum\limits_{{\bf q},{\bf q}_1}\langle\delta \hat{f}_1({\bf r},{\bf q},t)[\delta \hat{f}_1({\bf r},{\bf q}_1,t)+2\delta \hat{f}_2({\bf r},{\bf q}_1,t)  ]\rangle}{\big(\sum\limits_{\bf q} f_0({\bf r},{\bf q},t){+} f_1({\bf r},{\bf q},t)\big)^2},
 \end{equation}
 where the numerator and denominator are defined by Eqs. (\ref{22twnt1})-(\ref{36twnt7}), (\ref{26twnt}), (\ref{37p})-(\ref{39p}) and (\ref{32twnt5xx})-(\ref{34twnt6x}). Bringing together analytical and numerical calculations, we obtain $\sigma^2$ for different experimental conditions. Also, it is possible to compare the scintillation index obtained by employing different numbers  of iteration steps as described in this section.

 \begin{figure}[t!]
		\includegraphics[width=0.95\linewidth, keepaspectratio]{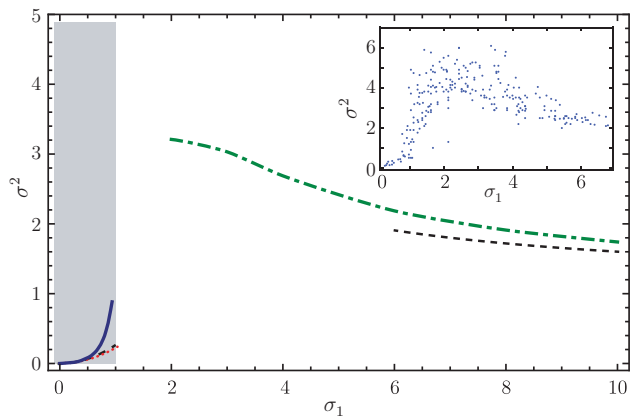}
\caption{(Color online) Scintillations as function of Rytov parameter. Results for different theoretical approaches qualitatively compared with experimental data. There are theoretical results of current paper (solid line) [Eq. (\ref{40p}) ], Rytov approach (dotted line) [Eq. (\ref{35twnt66}) ], asymptotic formulas for Huygens-Kirchhoff method  \cite{banakh79} (dashed lines) and the results of approach that the authors developed in Ref. \cite{enha} (dash-dotted line). The inset shows the typical experimental $\sigma^2$  for the considered atmospheric conditions (adopted from Ref. \cite{consortini} for $4\,\text{mm}<l_0\leq7\,\text{mm}$ ). Parameters for theories: $l_0=6.3\,$mm, $q_0=1.29\times10^{7}\,\text{m}^{-1}$, $r_0=0.01\,\text{m}$, $z=1200\,\text{m}$. The shaded area shows the parameter region considered in the current article.}

\label{fig:Experiment}
\end{figure}

\begin{figure}[ht]
			\includegraphics[width=0.97\linewidth,keepaspectratio]{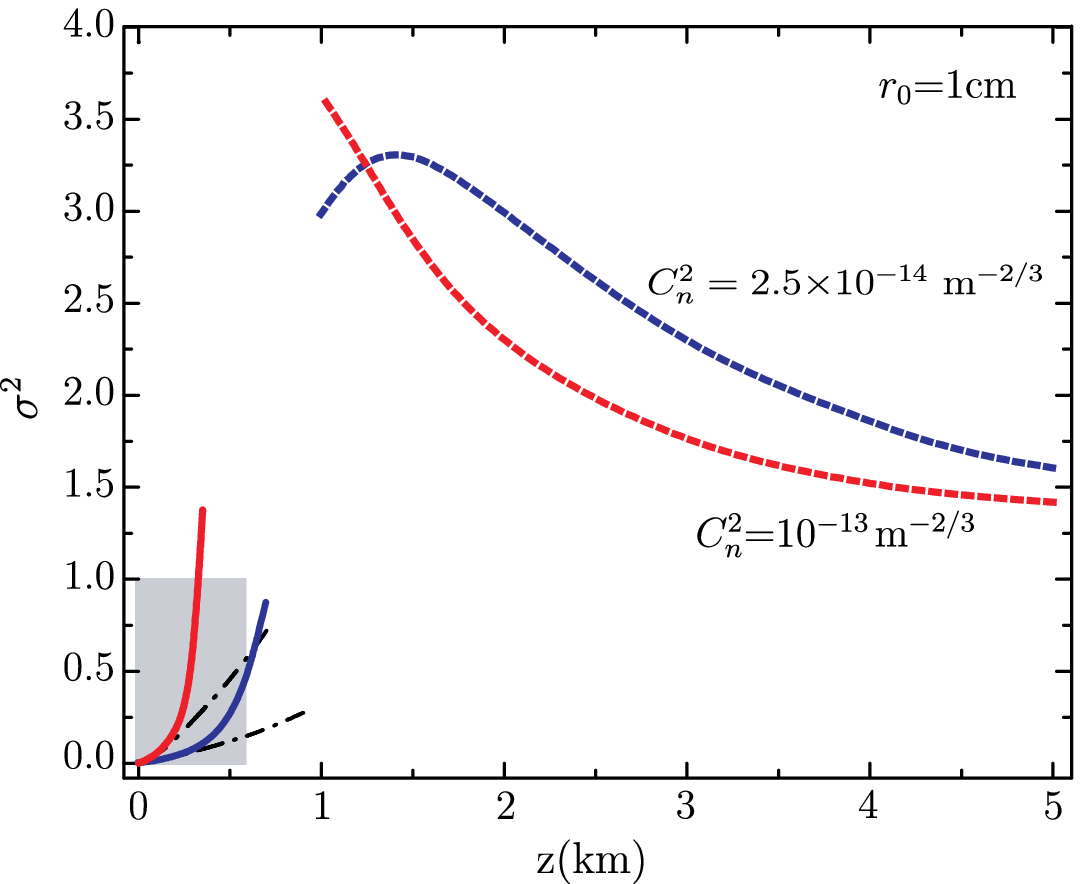}
			\includegraphics[width=0.97\linewidth,keepaspectratio]{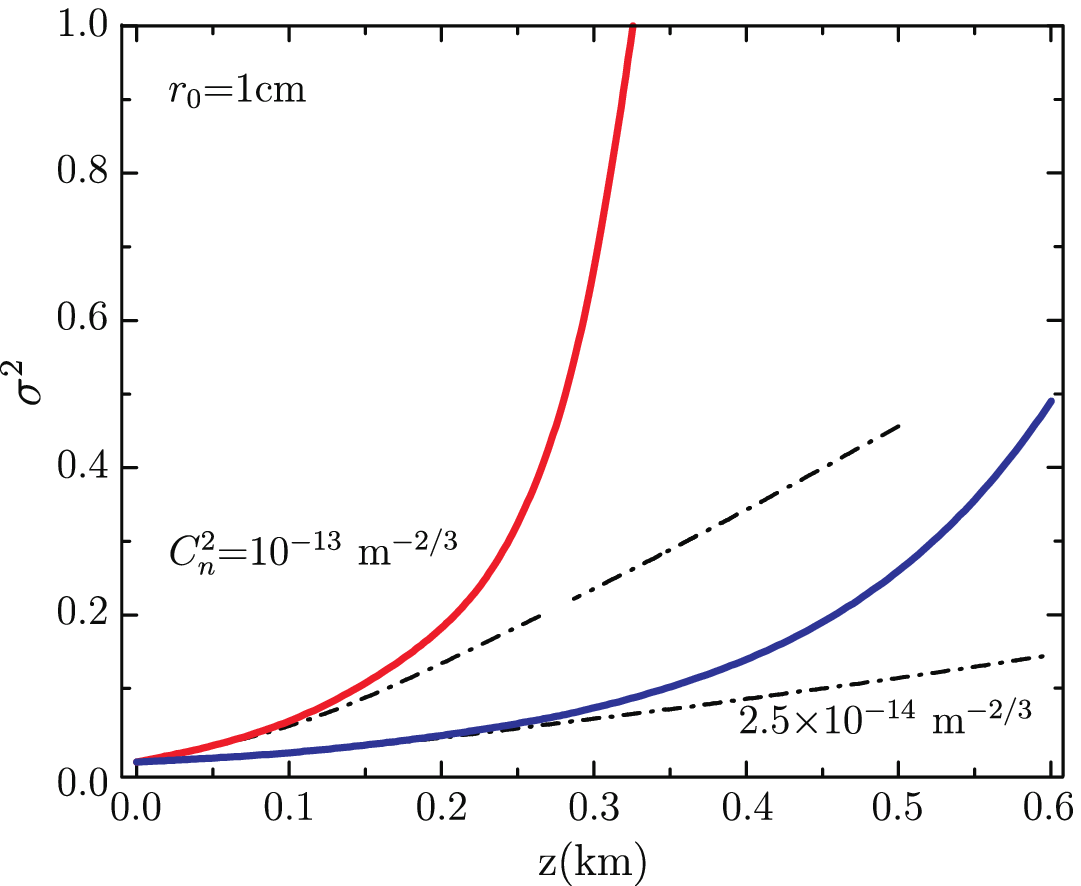}	
	\caption{(Color online) Scintillation index for coherent beams vs. propagation distance $z$. On the upper graph dash-dotted curves are obtained using the Rytov approach [Eq. (\ref{35twnt66})]; solid curves are obtained with the account for the collision term [Eq. (\ref{40p})]; dashed curves display the results obtained in Ref. \cite{enha} (see their Fig. 1 and 2), where the correlation of photon trajectories is accounted for. Shaded area at upper graph is enlarged and depicted on lower graph. Inner turbulence scale $\frac{l_0}{2\pi}=10^{-3}\,\text{m}$ and the optical wavelength $q_0=10^{7}\,\text{m}^{-1}$.\\}
	\label{fig:IntCollvsCrossCorell}
\end{figure}

 \section{Results and discussion}
\label{sec:discussion}

A complete theory of scintillations does not exist yet. At the same time, there are well-justified solutions in the limiting cases of weak ($\sigma_1^2\ll1$) and strong ($\sigma_1^2\gg1$) turbulences. The kinetic equation, in which a beam scattering is described by the collision integral, is applicable for any Rytov variance $\sigma_1^2$ with the exception of a very short distance equal to the typical eddy size. An exact solution of this equation is problematic.  Therefore, we restrict the numerical solution to a moderate values of $\sigma_1^2$ ($\sigma_1^2\leq 0.85$, see shaded area, in Fig. \ref{fig:Experiment}, and $\sigma_1^2\leq 0.75$ for the other figures) and use the iteration scheme described in Sec. \ref{sec:scint_ind}. At the same time this parameter is appreciably greater than  the range of the Rytov approach validity $\sigma_1^2<0.3$ \cite{Fante1}.

Figure \ref{fig:Experiment} compares the scintillation index calculated within the Boltzmann-Langevin approach with other theoretical approaches and with the typical experimental data, adopted from Consortini \textit{et al.} \cite{consortini}. Although the original data of Ref. \cite{consortini} are collected for spherical waves while theory deals with plane waves, we propose a qualitative comparison of results to illuminate peculiarities of scintillations and advantages of our method for their description. Naturally, for small values of the Rytov parameter ( $\sigma_1^2\leq 0.25$) our result coincides with the asymptotics for the Huygens-Kirchhoff method and Rytov-like method, but differs dramatically for the larger values showing the same increasing tendency as experimental data in a weak-to-moderate turbulence regime. For the sake of completeness we also provide theoretical results from the side of large values of Rytov parameter calculated within the approach of Ref. \cite{enha} and the Huygens-Kirchhoff approach. We observe that the Huygens-Kirchhoff method presents only a limited description for strong turbulence, while results of the approach from  Ref. \cite{enha} shows better description of scintillation index going deeper to the range of moderate turbulences. Moreover, the results of Ref. \cite{enha} show the tendency to mesh with the results of current paper plausibly repeating the overall behavior of scintillations in the cited experiment.
\begin{figure}
\includegraphics[width=0.97\linewidth,keepaspectratio]{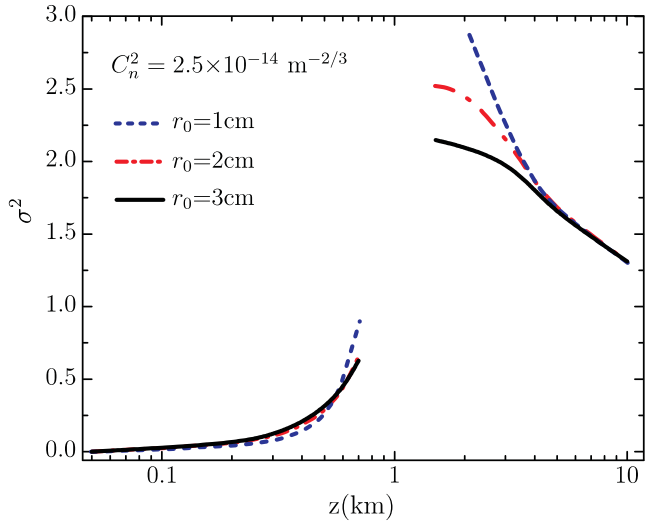}
\caption{(Color online) Scintillation index for coherent beam vs. propagation distance $z$ for different initial radii of the beam. The rest of the parameters are the same as in Fig. \ref{fig:IntCollvsCrossCorell}. The curves from the left side are obtained using the present  approach [Eq. (\ref{40p})]; the curves from the right side are obtained using the approach developed in Ref. \cite{enha}.}
\label{fig:DiffRadius}
\end{figure}

To take a closer look at our results we provide a comparison with the results of the Rytov approach under different configurations of atmospheric channel ( Fig. \ref{fig:IntCollvsCrossCorell}). Again for small values of $\sigma^2$, there is a good agreement for data obtained within the two approaches (enlarged shaded area at lower graph) and for greater values of $\sigma^2$ we can see not only numerical inconsistencies, but also different tendencies of $\sigma^2(z,C_n^2)$ to grow for considered cases. The comparison with the results of the previous paper \cite{enha} for moderate-to-strong turbulence regime displays the tendency for matching at some intermediate region. It also demonstrates that maximum of $\sigma^2$ should be situated at shorter distances $z$ if the structure constant, $C_n^2$, is larger. This result can be easily foreseen in view of the fact that strong photon-turbulence interaction approaches the crossover to the Gaussian statistics.

One more aspect taken under consideration is the dependence of scintillations on the initial radius of laser beam. Figure \ref{fig:DiffRadius} illustrates the behavior of the scintillation index in the regions adjoining the extremum of $\sigma^2$. We can see that the initial growth of $\sigma^2$ is steeper in the case of smaller initial radii $r_0$. This is due to stronger correlation of photon trajectories: the correlation is more pronounced for small $r_0$ \cite{enha}. This is easily explained since if the trajectories are closer to each other, then the probability for different photons to be scattered by the same eddy is greater. This is the case when a random scattering generates photon-photon correlations.

\begin{figure}
	\includegraphics[width=0.97\linewidth,keepaspectratio]{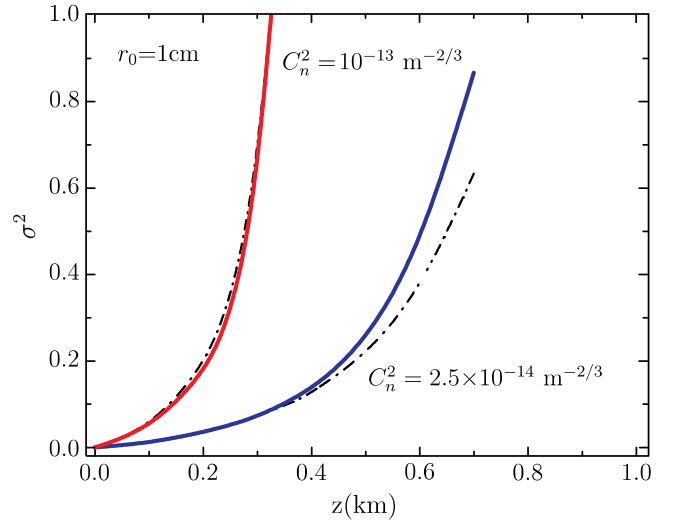}
	\caption{(Color online) Scintillation index as a function of propagation distance $z$ calculated with (solid) and without (dashed) accounting of contribution $\delta{\hat{f}_2}$. The parameters are the same as in Fig. \ref{fig:IntCollvsCrossCorell}.}
	\label{fig:popravka}
\end{figure}

Figure \ref{fig:popravka} can be used for explanation of the physical mechanism responsible for the increase of $\sigma^2$ in the range $\sigma_1^2\leq{0.75}$. The solid lines are obtained using Eq. (\ref{40p}). The data shown by the dash-dot line are obtained from the same expression considering $\delta \hat{f}_2=0$. There is only a small difference between the corresponding pairs of curves. Therefore, the major part of the discrepancy of our results for $\sigma^2$ from the results based on the Rytov approximation is due to the decrease of the photon density caused by the turbulence. This decrease is described by the term $f_1$ in the denominator of (\ref{40p}).

\section{Conclusion}
\label{sec:conclusion}

For decades, the description of light propagation in a turbulent atmosphere has remained a challenging theoretical problem.  The interconnection between the initial and the detected signals, obtained theoretically, is not sufficient for the description of  atmospheric communication system efficiency.  The point is that the detected signal has a memory about random scattering events occurred in the course of propagation. Therefore, even for the statistically homogeneous and stationary atmosphere, the received signal varies (fluctuates) for different paths. The size of these fluctuations is described by the scintillation index.

By definition, the scintillation index is expressed via the correlation functions of the photon distribution. The kinetic equation for the distribution function and its fluctuating part is derived here from first principles. Their solutions are obtained using the iteration procedure which is applicable for short propagation distances or small turbulence structure factors. In our analysis, we use the paraxial approximation for beams. This approximation reduces the problem to the case of a two-dimensional wave vector domain and simplifies the collision integral as well as correlation functions of the Langevin sources.

Concluding, we think that further progress in the problem of scintillations lies in the improvement of our ability to carry out complex multiple integrations.

\section{Acknowledgments}
The authors thank A. Gabovych, G. Berman, D. Vasylyev and E. Stolyarov for useful discussions and comments.

\appendix
\section{The collision integral}
\label{sec:appendix}
\numberwithin{equation}{section}
\setcounter{equation}{0}
The collision integral (\ref{12twelwe}) can be derived using the standard procedure. Nevertheless, some explanations are required.
The derivation of Boltzmann-like kinetic equations is based on the assumption of a negligibly short interaction time of individual particles (photons) with scatterers. \begin{widetext} The corresponding criteria are given by Eq. (\ref{9tenprime}). The other point concerns the explicit form of the scattering probability.  For our case, the collision process is described by the operator

\begin{equation}\label{A1}
\hat{J}=-i\frac{\omega _0}{V}\sum_{{\bf k},{\bf k'}}e^{-i{\bf k\cdot r}}n_{{\bf k}^\prime}
\big[b^\dag _{\bf q+ \frac {k}2}b_{\bf q-\frac{k}{2}+k^\prime}-b^\dag _{\bf q+ \frac {k}2-k^\prime}b_{\bf q-\frac{k}{2}} \big],
\end{equation}
(see Eq. (\ref{5seven})).
Using the quantity $b^\dag _{\bf q+ \frac {k}2}b_{\bf q-\frac{k}{2}+k^\prime}$, given by Eq. (\ref{8ten}), we rewrite Eq. (\ref{A1}) as
\begin{equation}\label{A2}
\hat{J}=-\hat{K}({\bf r},{\bf q},t)+ \frac{\omega_0^2}V\sum_{\bf k,k^\prime, {k}^{\prime\prime}}n_{\bf {k}^{\prime}}n_{\bf {k}^{\prime\prime}}\int\limits_{t_0}^{t}dt^\prime e^{-i{\bf k \cdot r}} \bigg[e^{i(\omega_{\bf q+\frac{k}{2}}-\omega_{\bf q-\frac{k}{2}+{k'}})\left(t-t'\right)}\big(b^\dag_{\bf q+\frac{k}{2}}b_{\bf q-\frac{k}{2}+{k'}+{k}^{\prime\prime}}-b^\dag_{\bf q+\frac{k}{2}-{k}^{\prime\prime}}b_{\bf q-\frac{k}{2}+{k'}}\big)
\end{equation}
\[ -e^{i(\omega_{\bf q+\frac{k}{2}-k^\prime}-\omega_{\bf q-\frac{k}{2}})\left(t-t'\right)}\big(b^\dag_{\bf q+\frac{k}{2}-k^\prime}b_{\bf q-\frac{k}{2}+{k}^{\prime\prime}}-b^\dag_{\bf q+\frac{k}{2}-k^\prime-{k}^{\prime\prime}}b_{\bf q-\frac{k}{2}}\big)\bigg]\bigg|_{t^\prime}=-\hat{K}({\bf r},{\bf q},t)+\hat{\tilde{J}},\]
where the second term in square brackets is derived from the first one by replacing ${\bf q} \rightarrow{\bf q-k^\prime}$ in the first one, and the interval $t-t_0$ satisfies the condition (\ref{9tenprime}).

Products of $n_{\bf {k}^{\prime}}n_{\bf {k}^{\prime\prime}}$ and $b^\dag b$ in Eq. (\ref{A2}) have a fluctuating nature. In what follows, we will neglect correlations between the corresponding subsystems. In this case we may consider them separately.

The quantity  $n_{\bf {k}^{\prime}}n_{\bf {k}^{\prime\prime}}$ contains a nonzero average constituent and  a fluctuating part. Let us consider the product $n_{\bf {k}^{\prime}}n_{\bf {k}^{\prime\prime}}$ in more details. By definition
\begin{equation}\label{A3}
n_{\bf {k}^{\prime}}n_{\bf {k}^{\prime\prime}}=\frac 1{V^2}\int\int d{\bf r}d{\bf r}_1e^{i[{\bf k}^\prime\cdot{\bf r}+{\bf k}^{\prime\prime}\cdot{\bf r}_1]}\delta n({\bf r})\delta n({\bf r}_1)=
\frac 1{V^2}\int\int d{\bf R}d{\bf s}e^{i({\bf k}^\prime+{\bf k}^{\prime\prime})\cdot{\bf R}+i({\bf k}^{\prime}- {\bf k}^{\prime\prime})\cdot{\bf s}/2}
\delta n({\bf R}+\frac {\bf s}2)\delta n({\bf R}-\frac {\bf s}2 ),
\end{equation}
where ${\bf R}=({\bf r}+{\bf r}_1)/2,\quad {\bf s}={\bf r}-{\bf r}_1$. The range of $s\lesssim l_{corr}$, where the correlation length $l_{corr}$ is comparable with the eddies size, provides a dominant contribution  into the average part of the integral (\ref{A3}).  In spatially homogeneous mediums, the quantity ${\langle\delta n({\bf R}{+}\frac {\bf s}2)\delta n({\bf R}{-}\frac {\bf s}2 )}\rangle$ does not depend on $\bf R$ and the characteristic values of $|{\bf k}^{\prime}- {\bf k}^{\prime\prime}|$ are restricted by $ 1/l_{corr}$.

The characteristic value of $R$ is of the order of the system size $L$. In this case
 $|{\bf k}^{\prime}+ {\bf k}^{\prime\prime}|{\sim}1/L$ tends to zero if $L{\rightarrow}\infty$. This means that the relation ${{\bf k}^{\prime}=-\bf k}^{\prime\prime}$ holds at any practically important values of ${\bf k}^{\prime}$ and ${\bf k}^{\prime\prime}$. Thus we have
 \begin{eqnarray}\label{A4}
\langle n_{\bf {k}^{\prime}}n_{\bf {k}^{\prime\prime}}\rangle &=& \frac 1V\delta_{{\bf k^\prime},-{\bf k^{\prime\prime}}}\int d{\bf s}\int \frac{d{\bf R}}Ve^{i{\bf k^\prime\cdot s}}\langle\delta n({\bf R}+\frac {\bf s}2)\delta n({\bf R}-\frac {\bf s}2 )\rangle\nonumber\\
&=&\delta_{{\bf k^\prime},-{\bf k^{\prime\prime}}}\int \frac {d{\bf s}}V e^{i{\bf k^\prime \cdot s}}\langle\delta n({\bf s})\delta n(0)\rangle\nonumber\\
&=&\delta_{{\bf k^\prime},-{\bf k^{\prime\prime}}}\langle n({\bf r})n(0)\rangle_{{\bf k}^\prime}=\delta_{{\bf k^\prime},-{\bf k^{\prime\prime}}}\langle |n_{\bf k^\prime}|^2  \rangle .
\end{eqnarray}
The angle brackets mean averaging over the volume $V$, which is assumed to be much greater than the correlation volume $l_{corr}^3$. Such averaging is equivalent to averaging over different configurations of turbulent atmosphere.

The substitution of $ \delta_{\bf {k}^\prime,-{\bf {k}}^{\prime\prime}}\langle|n_{\bf {k}^\prime}|^2\rangle$ for $n_{\bf {k}^{\prime}}n_{\bf {k}^{\prime\prime}}$ in Eq. (\ref{A2}) transforms the second term there to
\begin{eqnarray}\label{A5}
{\hat{\tilde{J}}=\frac{\omega_0^2}V\sum_{\bf k,k^\prime }\langle |n_{\bf k^\prime}|^2  \rangle\int_{t_0}^{t}dt^\prime e^{-i{\bf k\cdot r}} \bigg[e^{i(\omega_{\bf q+\frac{k}{2}}-\omega_{\bf q-\frac{k}{2}+{k'}})\left(t-t'\right)}\big(b^\dag_{\bf q+\frac{k}{2}}b_{\bf q-\frac{k}{2}}-b^\dag_{\bf q+\frac{k}{2}+{k}^{\prime}}b_{\bf q-\frac{k}{2}+{k'}}\big)}\nonumber\\
-e^{i(\omega_{\bf q+\frac{k}{2}-k^\prime}-\omega_{\bf q-\frac{k}{2}})\left(t-t'\right)}\big(b^\dag_{\bf q+\frac{k}{2}-k^\prime}b_{\bf q-\frac{k}{2}-{k}^{\prime}}-b^\dag_{\bf q+\frac{k}{2}}b_{\bf q-\frac{k}{2}}\big)\bigg]\bigg|_{t^\prime} .
\end{eqnarray}
The rest of the terms with $n_{\bf {k}^{\prime}}n_{\bf {k}^{\prime\prime}}$, where $\bf {k}^{\prime}\neq-\bf {k}^{\prime\prime}$, have a random nature and should be added to the Langevin source $\hat{K}({\bf r},{\bf q},t)$. These terms contribute negligibly to $\hat{K}$ and can be neglected if Eq. (\ref{9tenprime}) holds true.

For the short interval  $t-t_0$ [see (\ref{9tenprime})], the distribution function does not vary significantly and the evolution of operators $b^\dag_{\bf q+\frac{k}{2}}b_{\bf q-\frac{k}{2}}$ resembles the evolution in vacuum:
\begin{equation}\label{A6}
b^\dag_{\bf q+\frac{k}{2}}b_{\bf q-\frac{k}{2}}|_{t^\prime}=e^{-i(\omega_{\bf q+\frac{k}{2}}-\omega_{\bf q-\frac{k}{2}})\left(t-t'\right)}b^\dag_{\bf q+\frac{k}{2}}b_{\bf q-\frac{k}{2}}|_{t}.
\end{equation}
The operators in the right side of Eq. (\ref{A6}) depend only on a fixed time $t$ and the integration
in Eq. (\ref{A5}) concerns only the exponential functions
\begin{eqnarray}\label{A7}		
\int\limits_{t_0}^{t}dt^\prime e^{-i{\bf k\cdot r}} e^{i(\omega_{\bf q+\frac{k}{2}}-\omega_{\bf q-\frac{k}{2}+{k'}})\left(t-t'\right)}b^\dag_{\bf q+\frac{k}{2}}b_{\bf q-\frac{k}{2}}|_{t^\prime}
=b^\dag_{\bf q+\frac{k}{2}}b_{\bf q-\frac{k}{2}}|_t\int\limits_{t_0}^{t}dt^\prime e^{i(\omega_{\bf q-\frac{k}{2}}-\omega_{\bf q-\frac{k}{2}+{k'}})(t-t^\prime)}.
\end{eqnarray}
The condition (\ref{9tenprime}) enables the interval $t-t_0$ to be replaced by infinity

\begin{eqnarray}\label{A8}	
{\int\limits_{t_0}^{t}dt^\prime e^{i(\omega_{\bf q-\frac{k}{2}}-\omega_{\bf q-\frac{k}{2}+{k'}})(t-t^\prime)}
\approx\int\limits_0^{\infty}d\tau e^{i(\omega_{\bf q-\frac{k}{2}}-\omega_{\bf q-\frac{k}{2}+{k'}}+i\eta)\tau}}
=\frac i{\omega_{\bf q-\frac{k}{2}}-\omega_{\bf q-\frac{k}{2}+{k'}}+i\eta},
\end{eqnarray}
\end{widetext}
where $\eta\rightarrow +0$. Similar consideration is applicable to each term in Eq. (\ref{A5}). Then Eq. (\ref{A5}) reduces to

\begin{eqnarray}\label{A9}
\hat{\tilde{J}}&{=}&\frac{i\omega_0^2}V\sum_{\bf k,k^\prime }\langle |n_{\bf k^\prime}|^2  \rangle e^{-i{\bf k\cdot r}}\bigg[\frac {b^\dag_{\bf q+\frac{k}{2}}b_{\bf q-\frac{k}{2}}}{\omega_{\bf q-\frac{k}{2}}-\omega_{\bf q-\frac{k}{2}+{k'}}+i\eta}\nonumber\\
&&-\frac{b^\dag_{\bf q+\frac{k}{2}+{k}^{\prime}}b_{\bf q-\frac{k}{2}+{k'}}}{\omega_{\bf q+\frac{k}{2}}-\omega_{\bf q+\frac{k}{2}+{k'}}+i\eta}-
\frac {b^\dag_{\bf q+\frac{k}{2}+k^\prime}b_{\bf q-\frac{k}{2}+k^\prime}}{\omega_{\bf q-\frac{k}{2}+k^\prime}-\omega_{\bf q-\frac{k}{2}}+i\eta}\nonumber\\
&&+\frac{b^\dag_{\bf q+\frac{k}{2}}b_{\bf q-\frac{k}{2}}}{\omega_{\bf q+\frac{k}{2}+k^\prime}-\omega_{\bf q+\frac{k}{2}}+i\eta}\bigg]\bigg|_t.
\end{eqnarray}

In the last two terms, the value of  ${\bf k^\prime}$ is  replaced by ${-\bf k^\prime}$. For paraxial beams, considered here, we can use the approximation  $\omega_{\bf q}=cq\approx cq_z$, which implies a negligible contribution of $q_{x,y}$ components. Then, using the relation
\[\frac 1{ck^\prime_z-i\eta} -\frac 1{ck^\prime_z+i\eta}=\frac{2\pi i }c\delta(k^\prime_z)\]
and integration over $k^\prime_z$, Eq. (\ref{A9}) simplifies to
 \begin{equation}\label{A10}
  \hat{\tilde{J}}=\frac{2\pi\omega_{0}^{2}}{c}\int d{\bf k'_{\bot}}\psi({\bf k'_{\bot}})\big(\hat{f}({\bf r},{\bf q},t)-\hat{f}({\bf r},{\bf q+k'_{\bot}},t)\big),
 \end{equation}
 where the definition (\ref{1threee}) of PDF  was used. Equation (\ref{A10}) coincides with the collision integral $\hat{\nu}_{\bf q}\big \{ \hat{f}({\bf r},{\bf q},t)\}$ represented by Eq. (\ref{12twelwe}).


\section{Boundary conditions for the incident light}
\label{sec:appendix1}

Calculation of concrete parameters of laser radiation is possible if the boundary conditions for the incident light are specified. Usually, the Gaussian distribution of the laser field in the aperture plane is assumed
\begin{equation}\label{30twnt4}
\Phi({\bf r}_\bot)=(2/\pi r_0^2)^{1/2}e^{-{r^2_\bot}/{r^2_0}},
\end{equation}
where $r_0$ is the aperture radius.
The laser and outgoing field should match in the aperture plane. This means that
\begin{equation}\label{31twnt5x}
\sum_{{\bf q}_{\bot},q_z}\bigg( \frac{2\pi\hbar\omega_{\bf q}}V\bigg)^{1/2}b_{\bf q}e^{-i\omega_{\bf q}t+i{\bf q}_{\bot}\cdot{\bf r}_{\bot}}=\alpha_Lb\Phi({\bf r}_{\bot})e^{-i\omega_0t},
\end{equation}
where $b$ is the amplitude of the laser mode, and the coefficient $\alpha_L$ describes penetration of this field through the aperture. As before, the paraxial approximation ($\omega_{\bf q}\approx cq_z$) can be used. Also, the requirement of synchronism of both fields, restricts the  left-hand side sum with terms $q_z=\omega_0/c=q_0$.
Then the explicit value for $b_{\bf q}$ follows from Eq. (\ref{31twnt5x})
\begin{equation}\label{32twnt5xx}
b_{\bf q}=b\alpha_L\frac {r_0}{\sqrt{\hbar \omega_0}}\sqrt{\frac{L_z}S}e^{-q^2_{\bot}r_0^2/4}\delta_{q_z,q_0},
\end{equation}
 which determines the boundary value of PDF:
\begin{equation}\label{33twnt5}
f({\bf r_\bot},z{=}0,{\bf q},t)=\delta_ {q_z,q_0}b^\dag(t) b(t)\frac {2{\alpha_L}^2}{\pi S\hbar\omega_0}e^{-q_{\bot}^2r_0^2/2-{2r_{\bot}^2}/{r_0^2}}.
\end{equation}

The extension of Eq. (\ref{33twnt5}) for the case of a partially coherent beam is realized by substituting $\frac{q_\bot^2r_1 ^2}2$ for $\frac{q_\bot^2r_0^2}2$  \cite{Chu}. Here  $r^2_1=r_0^2/(1+2r_0 ^2\lambda _c^{-2})$, and the quantity $\lambda _c$ describes the effect of the phase diffuser which is used for suppression of scintillations. The mentioned modification of the initial distribution expands the range of  $q_\bot $ variation to the values of the order $\frac{\pi}r_1 $ and does not affect the spatial distribution in the ${\bf r}_\bot$-domain. The diffuser influence vanishes in the limit of  $\lambda _c\rightarrow\infty$ because in this case ${ r}_1\rightarrow r_0$.

In the case of ${\bf r}_\bot=0$, the denominator in Eq. (\ref{22twnt1}) is given by
\begin{equation}\label{34twnt6x}
 \sum_{\bf q} f_0({\bf r},{\bf q},t)=\frac {\alpha_L^2r_1^2q_0\langle b^\dag b\rangle}{\pi^2\hbar c(4+\rho_0^2\rho_1^2)},
\end{equation}
where the derivation of (\ref{34twnt6x}) was somewhat simplified by inserting $L_0^{-1}=0$ in Eq. (\ref{13twelwwe}), $\rho _{0,1}^2={r_{0,1}^2q_0}/z $.

\end{document}